\pdfoutput=1
\documentclass[prl,twocolumn,showpacs,amsmath,amssymb,superscriptaddress,floatfix]{revtex4}

\usepackage{graphicx}
\usepackage{dcolumn}
\usepackage{bm}

\begin{document}

\title{Quantum coherence and its dephasing in the giant spin Hall effect and nonlocal voltage generated by magnetotransport through multiterminal graphene bars}

\author{Chien-Liang Chen}
\affiliation{Department of Physics and Astronomy, University of Delaware, Newark, DE 19716-2570, USA}
\affiliation{Department of Physics, National Taiwan University, Taipei 10617, Taiwan}
\author{Ching-Ray Chang}
\email{crchang@phys.ntu.edu.tw}
\affiliation{Department of Physics, National Taiwan University, Taipei 10617, Taiwan}
\author{Branislav K. Nikoli\' c}
\email{bnikolic@udel.edu}
\affiliation{Department of Physics and Astronomy, University of Delaware, Newark, DE 19716-2570, USA}

\begin{abstract}
Motivated by the recent experimental observation [D. A. Abanin {\em et al.}, Science {\bf 323}, 328 (2011)] of nonlocality in magnetotransport near 
the Dirac point in six-terminal graphene Hall bars, for a wide range of temperatures and magnetic fields, we develop a nonequilibrium Green function (NEGF) theory of this phenomenon. In the phase-coherent regime and strong magnetic field, we find large spin Hall (SH) conductance in four-terminal bridges, where the SH current is pure only at the Dirac point (DP), as well as the nonlocal voltage at a remote location in six-terminal bars where the direct and inverse SH effect operate at the same time. The ``momentum-relaxing'' dephasing reduces their values at the DP by two orders of magnitude while concurrently washing out any features away from the DP. Our theory is based on the Meir-Wingreen formula with dephasing introduced via phenomenological many-body self-energies, which is then linearized for multiterminal  geometries to extract currents and voltages.
\end{abstract}

\pacs{72.80.Vp,72.15.Gd,72.10.Bg}
\maketitle

The recent experiments~\cite{Abanin2011} on magnetotransport near the Dirac point (DP) in graphene have unveiled yet another exotic electronic property of this one-atom-thick carbon allotrope~\cite{Geim2009} which involves nonlocality and quantum mechanics while manifesting even at room temperature. The traditional observation of nonlocality, where current is injected through a pair of terminals and voltage is measured between another pair of terminals at some {\em remote} location, requires two-dimensional (2D) systems placed in a strong external magnetic field to generate the integer quantum Hall effect~\cite{Qi2011} (QHE) or spin-orbit coupling~\cite{Winkler2003} (SOC) that can give rise to  mesoscopic~\cite{Nikolic2005b,Brune2010} or quantum~\cite{Qi2011,Roth2009} spin Hall effects (SHEs). In the former case, nonlocality is due to transport through chiral edge states, while in the latter case injected longitudinal charge current generates transverse spin Hall current which is then detected in the remote part of the device via the inverse-SHE-induced voltages~\cite{Brune2010,Hankiewicz2004} on the proviso that spins can survive dephasing between two locations.

On the other hand, nonlocal voltage was observed in Ref.~\cite{Abanin2011} even in weak magnetic fields $B \simeq 1$ T and at room temperature $T=300$ K  which is outside of the integer QHE regime. Also, high mobility graphene samples were supported by substrate made of atomically flat hexagonal boron-nitride that rules out Rashba SOC~\cite{Winkler2003}, introduced by charge impurities from the substrate~\cite{Ertler2009} or lattice distortion by adatoms~\cite{CastroNeto2009}, that would be responsible for the mesoscopic SHE scenario~\cite{Nikolic2005b}.

\begin{figure}
\includegraphics[scale=0.35,angle=0]{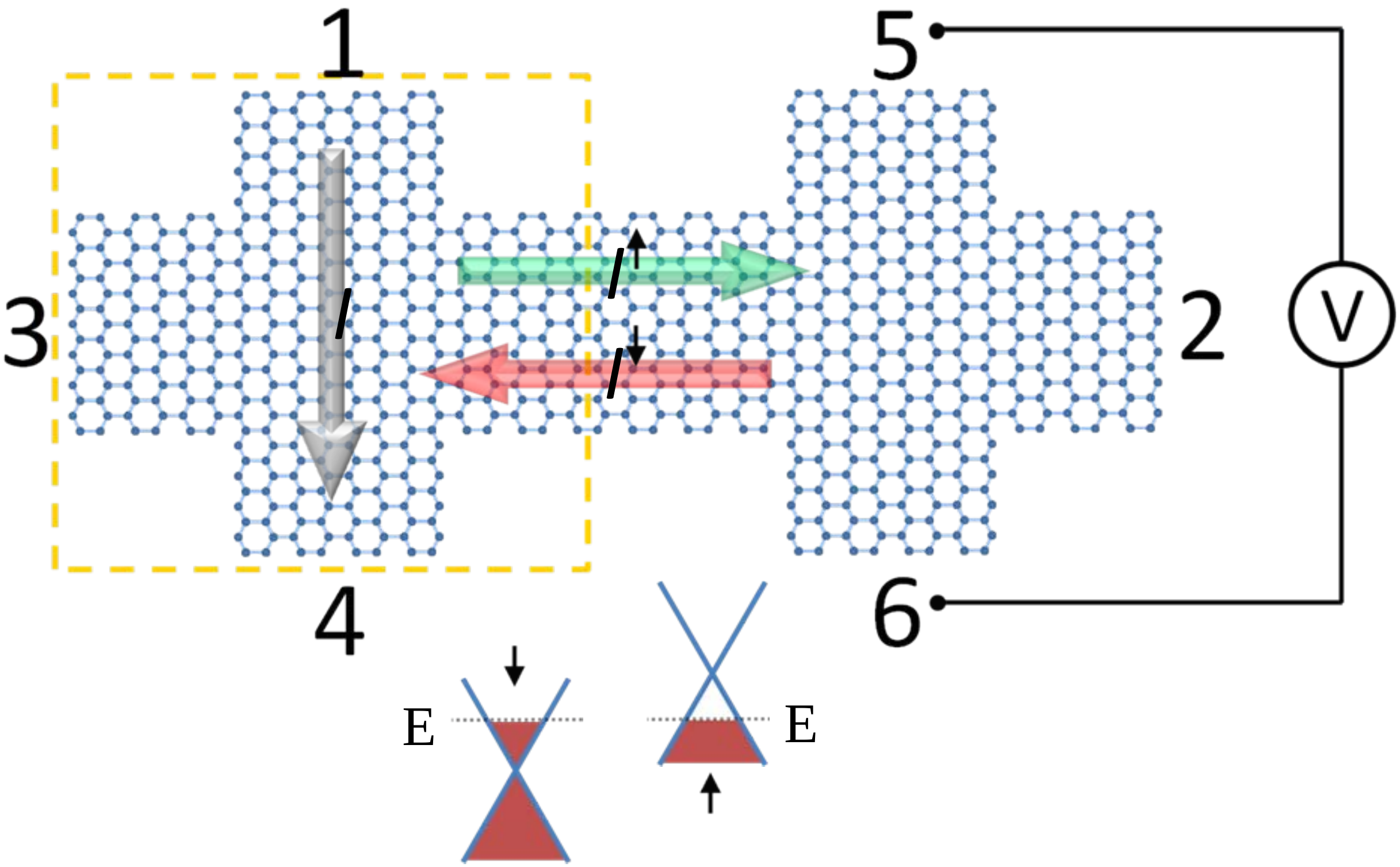}
\caption{(Color online) Schematic view of the six-terminal graphene Hall bar, modeled on the tight-binding lattice with single $\pi$ orbital per site, which is employed to investigate nonlocal voltage between leads 5 and 6 due to current injected between leads 1 and 4. The dashed box on the left marks the four-terminal bridge used in the analysis of Zeeman-splitting-driven SHE where current injected between leads 1 and 4 induces spin current in leads 2 and 3. The active region of the bar consists of graphene nanoribbon with armchair edges (AGNR) and a portion of semi-infinite leads modeled as GNRs with zigzag edges. For simplicity, external magnetic field, or many-body interactions responsible for dephasing, are present only within the illustrated active region.}
\label{fig:fig1}
\end{figure}

It turns out that SHE in the absence of SOC  has a simple intuitive explanation based on classical Newtonian dynamics of massless Dirac fermions. The classical Hamiltonian of low-energy quasiparticles close to the Dirac point (DP) is given by $H^{\pm }({\bm p})=\pm v_{F}\sqrt{p_{x}^2+p_{y}^2}$, which in the weak  external magnetic field ${\bf B}=\nabla \times {\bf A}$ becomes $H^{\pm}({\bm p})=\pm v_{F}\sqrt{(p_{x}-eA_{x})^2+(p_{y}-eA_{y})^2}$. The classical velocity is then given by $v_x^\pm = \partial H^{\pm}/\partial p_x=\pm v_{F} {\bm \Pi}_x/\sqrt{{\bm \Pi}^2}$, where ${\bm \Pi}={\bm p} - e{\bf A}$, and the corresponding acceleration is
\begin{equation}
{\bm a}^\pm = \frac{d {\bm v}}{d t} = \frac{e v_{F} {\bm v}^\pm \times {\bf B}}{\sqrt{{\bm \Pi}^2}} = \pm \frac{e v^2_{F} {\bm v}^\pm \times {\bf B}}{E^\pm}.
\end{equation}
Thus, the quasiparticles with energy $E^+$ above the DP (or below with energy $E^-$) moving in a weak (i.e., non-quantizing) perpendicular magnetic field will experience a transverse force which deflects them to the left (right). Furthermore, when $E^\pm$ is very close to the DP such deflecting force will be very large. Although the Zeeman splitting $\Delta_{\rm Z}$ in 2D electron gases (2DEGs) is typically small in a weak external magnetic field~\cite{Winkler2003}, it can play an essential role in graphene for $k_BT < \Delta_{\rm Z}$ by shifting the Dirac cones for opposite spins to induce two types of carriers illustrated in the lower inset in Fig.~\ref{fig:fig1}. The quasiparticles with energy $E^+$ are spin-$\uparrow$ polarized while those with energy $E^-$ are spin-$\downarrow$ polarized. These two effects, classical for charge and quantum for spin, conspire to generate transverse spin current in response to longitudinal charge current, as illustrated in Fig.~\ref{fig:fig1}. Such phenomenology is similar to SHE in 2DEG bridges~\cite{Nikolic2005b,Brune2010}, even though no SOC is involved to provide the deflecting force of opposite direction for opposite  spins~\cite{Nikolic2005c}.

These simple arguments for the existence of the Zeeman-splitting-driven SHE (ZSHE) in graphene  can be converted into a quasiclassical transport theory based on the Boltzmann equation~\cite{Abanin2011a}. However, quasiclassical theory is valid in high-$T$ and weak-$B$ regime, while experiments~\cite{Abanin2011} have observed increasingly more profound nonlocality into the low-$T$ and/or high-$B$ regime so that a unified theory is called for that can cover such wide range of parameters. For example, such theory should explain the nonlocal voltage in strong (quantizing) external magnetic field but at intermediate temperatures where edge-state transport mechanism is removed.

\begin{figure}
\includegraphics[scale=0.32,angle=0]{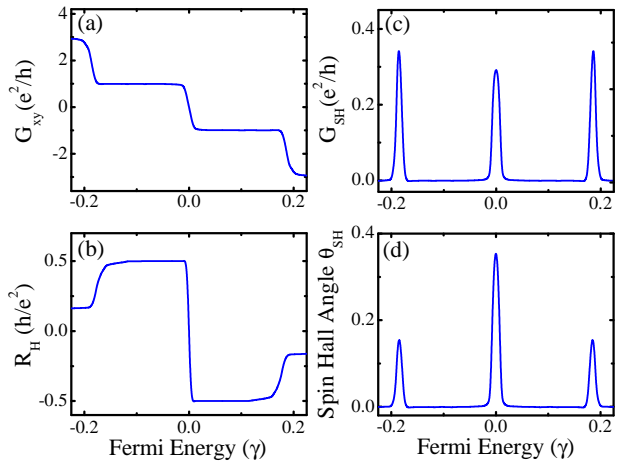}
\caption{(Color online) The charge and spin transport quantities in the {\em four-terminal} graphene bridge: (a) charge Hall conductance $G_{xy}=I_2/(V_1-V_4)$; (b) charge Hall resistance $R_{\rm H}=(V_3 - V_2)/I_{1}$; (c) spin Hall conductance $G_{\rm SH}=I_2^S/(V_1-V_4)$; and (d) spin Hall angle $\theta_{\rm SH}=I_2^S/I_1$. The width of AGNR channel is $W/\ell_B=3.42$ in the units of the magnetic length $\ell_B$ and a small ``momentum-relaxing'' dephasing $d_m=0.04 \gamma$ in introduced into the active region shown in Fig.~\ref{fig:fig1}.}
\label{fig:fig2}
\end{figure}
\begin{figure}
\includegraphics[scale=0.32,angle=0]{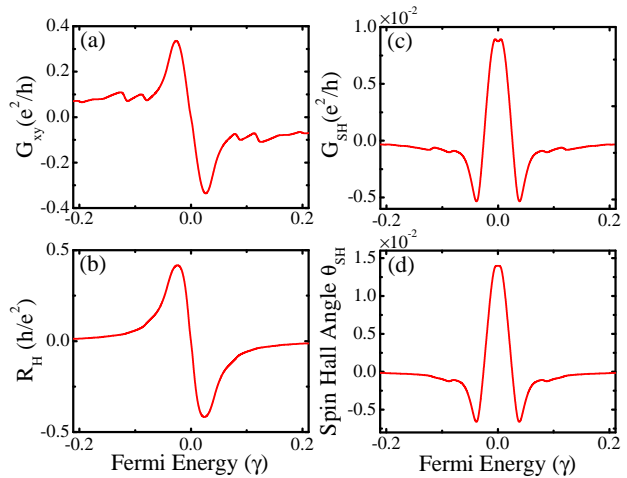}
\caption{(Color online) The charge and spin transport quantities in the {\em four-terminal} graphene bridge: (a) charge Hall conductance $G_{xy}$; (b) charge Hall resistance $R_{\rm H}$; (c) spin Hall conductance $G_{\rm SH}$; and (d) spin Hall angle $\theta_{\rm SH}$. The width of the AGNR channel is $W/\ell_B=1.53$ in the units of the magnetic length $\ell_B$ and large ``momentum-relaxing'' dephasing $d_m=0.4 \gamma$ in introduced into the active region shown in Fig.~\ref{fig:fig1}.}
\label{fig:fig3}
\end{figure}

\begin{figure}
\includegraphics[scale=0.32,angle=0]{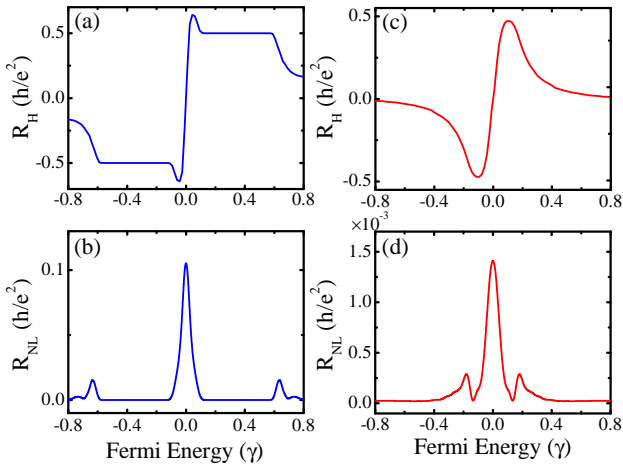}
\caption{(Color online) Panels (a) and (c) plot charge Hall resistance $R_{\rm H} = (V_1 - V_4)/I_{3}$, while panels (b) and (d) plot {\em nonlocal} resistance $R_{\rm NL} = (V_5 - V_6)/I_{1}$ as the central quantity measured in the recent experiments~\cite{Abanin2011} on {\em six-terminal} graphene Hall bars. The quantum coherence is retained in panels (a) and (b) where only a small ``momentum-relaxing'' dephasing $d_m=0.02\gamma$ is present in the active region of the bar, while much larger  dephasing $d_m=0.5\gamma$ is used for panels (c) and (d). The width of the AGNR channel in Fig.~\ref{fig:fig1} is $W/\ell_B=3.42$ in panels (a) and (b) and $W/\ell_B=1.53$ in panels (c) and (d) in the units of the magnetic length $\ell_B$.}
\label{fig:fig4}
\end{figure}

In this Letter, we develop a {\em fully quantum transport theory} of ZSHE in four-terminal graphene bridges, illustrated by the device within a dashed box in Fig.~\ref{fig:fig1}, as well as the nonlocal voltage induced by the combination of direct and inverse ZSHE in six-terminal Hall bars shown in Fig.~\ref{fig:fig1}. This approach intrinsically accounts for the contributions of both electrons and holes, which is crucial to describe transport near the DP~\cite{Abanin2011a}, and it can also handle arbitrary scattering processes (in contrast to the Boltzmann equation which breaks   down~\cite{Klos2010} close to the DP). Our central results, summarized in Figs.~\ref{fig:fig2}, \ref{fig:fig3} and \ref{fig:fig4}, interpolate smoothly between the phase-coherent transport regime at low-$T$ in the quantizing external magnetic field and the semiclassical transport regime where dephasing by many-body interactions destroys features found at low-$T$ while leaving peaks (of reduced magnitude though) in the SH conductance and nonlocal voltage around the DP in accord with experimental observations~\cite{Abanin2011}.

 In the analysis of four-terminal bridges, voltage $V/2$ is applied to lead 1 and $-V/2$ to lead 4 while voltages on leads 2 and 3 are set to zero. Figure~\ref{fig:fig2} shows that in the quantizing external magnetic field $W/\ell_B>1$, where $\ell_B=\sqrt{\hbar/|eB|}$ is the magnetic length in graphene, the four-terminal bridge generates large spin Hall conductance $G_{\rm SH} = (I_2^\uparrow - I_2^\downarrow)/(V_1-V_4)$ and the corresponding SH angle \mbox{$\theta_{\rm SH} = G_{\rm SH}/G_L$} with $G_L=I_1/(V_1-V_4)$ being the longitudinal charge conductance. The spin current $I_2^S=I_2^\uparrow - I_2^\downarrow$ in the ZSHE is carried by spins polarized along the \mbox{$z$-axis} orthogonal to the plane of graphene.  The value of $G_{\rm SH}$ is comparable to the one predicted~\cite{Nikolic2009} for the Rashba SO-coupled four-terminal 2DEGs of the size of the spin precession length (on which spin-$\uparrow$ precesses to spin-$\downarrow$ state). Unlike mesoscopic SHE~\cite{Nikolic2005b} in 2DEGs where Rashba SOC induces both the transverse spin deflection~\cite{Nikolic2005c} and spin dephasing which compete against each other in the processes of generating {\em pure} (not accompanied by any net charge flux) spin current, in the ZSHE transverse spin current is pure only at the DP [where charge current $I_2=I_2^\uparrow + I_2^\downarrow$ becomes zero in Figs.~\ref{fig:fig2}(a) and ~\ref{fig:fig3}(a)] and spin precession is {\em absent}. This might be advantageous for spintronic applications since spin dephasing is evaded, as demonstrated by the experimental detection of nonlocal voltage even at distances $\sim 10$ $\mu$m away from the device region where SH current was induced~\cite{Abanin2011}. We  note that for very strong magnetic field, as could be achieved in ferromagnetic graphene, the $G_{\rm SH}$ peaks in Fig.\ref{fig:fig2}(c) would become quantized~\cite{Sun2010} as a realization of quantum SHE~\cite{Qi2011} in the absence of SOC.

The introduction of dephasing processes into the four-terminal bridge, which relax {\em both}~\cite{Golizadeh-Mojarad2007a} the phase and the momentum of quasiparticles propagating through the active region, destroys the quantization of the charge Hall conductance $G_{xy}=I_2/(V_1-V_4)$ or charge Hall resistance  $R_{\rm H}$ and underlying chiral edge states, as demonstrated by the transition from Fig.~\ref{fig:fig2}(a) to Fig.~\ref{fig:fig3}(a) for $G_{xy}$ and from Fig.~\ref{fig:fig2}(b) to Fig.~\ref{fig:fig3}(b) for $R_{\rm H}$. The charge Hall resistance in four-terminal bridges is defined as $R_{\rm H} = (V_3-V_2)/I_1$ for the measuring setup where current $I_1$ is injected into lead 1 and voltages $V_3$ and $V_2$ develop as the response to it. The SH conductance and SH angle are concurrently reduced by two orders of magnitude, which are values similar to those found in quasiclassical approaches~\cite{Abanin2011a} in the temperature range \mbox{$T= \text{200--300 K}$}.

 In the analysis of six-terminal Hall bars,  charge current $I_1$ is injected through lead 1 and current $-I_1$ flows through lead $4$ while $I_\alpha \equiv 0$ in all other leads. We then compute voltages which develop in the leads $\alpha=2,3,5,6$ labeled in Fig.~\ref{fig:fig1} in response to injected current $I_1$. Figure~\ref{fig:fig4}(b) shows peaks in the nonlocal resistance, defined as $R_{\rm NL} = (V_5-V_6)/I_{1}$, within the phase-coherent transport regime which closely resemble the DP and side peaks observed experimentally in strong (quantizing) external magnetic field~\cite{Abanin2011}. The transition of $R_{\rm NL}$ from Fig.~\ref{fig:fig4}(b) to Fig.~\ref{fig:fig4}(d) shows how dephasing removes both side peaks while leaving the nonlocal voltage  around the DP which is two orders of magnitude smaller than in the phase-coherent regime. The Hall resistance in the six-terminal bar, $R_{\rm H} = (V_1-V_4)/I_3$ defined for current injected $I_3$ and voltages measured between leads 1 and 4 (for $I_1=I_4=0$), changes smoothly from Fig.~\ref{fig:fig4}(a) to Fig.~\ref{fig:fig4}(c) as dephasing in increased, where the curve in Fig.~\ref{fig:fig4}(c) looks exactly the same as those observed experimentally for \mbox{$T=250$ K} and $B= \text{1--12 T}$~\cite{Abanin2011a}.

In the rest of the paper, we discuss details of the microscopic Hamiltonian model for multiterminal graphene Hall bar in Fig.~\ref{fig:fig1} and the nonequilibrium Green function (NEGF) formalism including dephasing which is applied to understand transport properties of this device. Close to the DP, graphene can be can be described by the tight-binding Hamiltonian with single $\pi$-orbital per site
\begin{equation}\label{eq:hamilton}
\hat{H} = \sum_{\bf n} (\varepsilon_{\bf n} + g \mu_B \sigma B )\hat{c}_{{\bf n} \sigma}^\dagger \hat{c}_{{\bf n} \sigma} - \gamma \sum_{\langle {\bf nm} \rangle, \sigma} e^{i \phi_{{\bf nm}}} \hat{c}_{{\bf n} \sigma}^\dagger \hat{c}_{{\bf m} \sigma}.
\end{equation}
Here $\varepsilon_{\bf n}$ is the on-site energy,  $\sigma = +1$ for spin-$\uparrow$ electron and $\sigma = -1$ for spin-$\downarrow$ electron so that Zeeman splitting is given by $\Delta_{\rm Z}=2 g \mu_B B$, $\hat{c}_{{\bf n} \sigma}^\dag$ ($\hat{c}_{{\bf n} \sigma}$) creates (annihilates) electron with spin $\sigma$ in the $\pi$-orbital located on site ${\bf n}$, and $\gamma$ is the nearest-neighbor hopping parameter. The external magnetic field enters through the phase factor $\phi_{\bf nm}=\frac{2\pi}{\phi_0} \int_{\bf n}^{\bf m} \mathbf{A} \cdot d{\bf s}$ where the vector potential \mbox{$\mathbf{A}=(By,0,0)$} is chosen in the Landau gauge and $\phi_0=h/e$ is the flux quantum. The weak vs. strong magnetic field is tuned using the ratio $W/\ell_B$, where $W$ is the width of the AGNR channel of the bar in Fig.~\ref{fig:fig1}. All graphene bars studied in Figs.~\ref{fig:fig2}, ~\ref{fig:fig3} and ~\ref{fig:fig4} are placed in the quantizing magnetic field, $W/\ell_B>1$.

While the integer QHE and quantum SHE have introduced the intricate physics of topologically ordered phases~\cite{Qi2011}, their operational description used to analyze transport measurements~\cite{Brune2010,Roth2009} is typically based on the multiterminal Landauer-B\"{u}ttiker (LB)  formulas~\cite{Buttiker1986}
\begin{equation}\label{eq:mlb}
I_\alpha = \frac{e^2}{h} \sum_\beta T_{\alpha \beta}(V_\alpha - V_\beta),
\end{equation}
written here assuming zero temperature. They relate charge current in lead $\alpha$ to voltages $V_\beta$ in all other leads attached to the sample via the matrix of transmission coefficients $T_{\alpha \beta}$. These formulas are valid when phase coherence is retained in the active region of the device, while phase breaking events are assumed to be taking place only in the reservoirs to which the leads are attached at infinity and where electrons are equilibrated to acquire the Fermi-Dirac distribution $f_\alpha(E)=f(E-eV_\alpha)$.

To take into account dephasing effects phenomenologically, B\"{u}ttiker introduced~\cite{Buttiker1985} an elegant concept of voltage probes attached to the active region where no net current flows through them, so that for every electron that enters the probe and is absorbed by its reservoir another one has to come out which is not coherent with the one going in. For example, to apply this method to the graphene Hall bar in Fig.~\ref{fig:fig1}, one can attach one-dimensional leads to each site~\cite{Metalidis2006} of the honeycomb lattice. This is equivalent to adding a complex energy $-i\eta$ to $\varepsilon_{\bf n}$ in the Hamiltonian Eq.~\eqref{eq:hamilton} (parameter $\eta$ is related to the dephasing time $\eta=\hbar/2\tau_\phi$). In addition, one has to solve Eq.~\eqref{eq:mlb} by imposing that current through extra 1D leads is zero thereby ensuring conservation of the total charge current.

However, besides washing out quantum-coherence-generated fluctuations in $T_{\alpha \beta}$, B\"uttiker voltage probes are also introducing additional scattering  (i.e., reduction of $T_{\alpha \beta}$) in an uncontrolled fashion~\cite{Golizadeh-Mojarad2007a}. The NEGF formalism~\cite{Haug2007} provides a rigorous prescription for including any dephasing process to any order by starting from a microscopic
Hamiltonian and by constructing interacting self-energies due to electron-electron, electron-phonon~\cite{Haug2007} or electron-spin~\cite{Hurley2011} interactions. Although the NEGF formalism is virtually the only fully quantitative quantum transport approach capable of scaling to large systems~\cite{Areshkin2010}, the self-consistent computation of self-energies by starting from some microscopic many-body Hamiltonian is at present prohibitively expensive for devices containing realistic number of atoms. Thus, to include dephasing processes in the device in Fig.~\ref{fig:fig1} containing few thousands of carbon atoms, we adopt a phenomenological model of Ref.~\cite{Golizadeh-Mojarad2007a} that is comparable to B\"{u}ttiker voltage probes in conceptual and numerical simplicity, and yet allows one the flexibility of adjusting the degree of phase and momentum relaxation independently.

The two fundamental objects~\cite{Haug2007} of the NEGF formalism are the retarded  \mbox{$G^{r,\sigma\sigma'}_{\bf nm}(t,t')=-i \Theta(t-t') \langle \{\hat{c}_{{\bf n}\sigma}(t) , \hat{c}^\dagger_{{\bf m}\sigma'}(t')\}\rangle$} and the lesser \mbox{$G^{<,\sigma\sigma'}_{\bf nm}(t,t')=i \langle \hat{c}^\dagger_{{\bf n}\sigma'}(t') \hat{c}_{{\bf m} \sigma}(t)\rangle$} GFs which describe the density of available quantum states and how electrons occupy those states, respectively. Here $\langle \ldots \rangle$ denotes the nonequilibrium statistical average~\cite{Haug2007}. In stationary problems, $\hat{G}^r$ and $\hat{G}^<$ depend only on the time difference $t-t^\prime$ or energy $E$ after Fourier transformation. Their matrix representations in the basis of local orbitals, such as the $\pi$ ones in Eq.~\eqref{eq:hamilton}, satisfy the following equations
\begin{eqnarray}\label{eq:negf}
{\bf G}^r(E) & = & \left[E - {\bf H} - \sum_\alpha {\bm \Sigma}^r_\alpha(E-eV_\alpha) - {\bm \Sigma}_{\rm int}^r(E) \right]^{-1}, \label{eq:negf1} \\
{\bf G}^<(E) & = & {\bf G}^r(E) \left[\sum_\alpha {\bm \Sigma}^<_\alpha(E) + {\bm \Sigma}^<_{\rm int}(E)\right] {\bf G}^a(E). \label{eq:negf2}
\end{eqnarray}
Here ${\bm \Sigma}^r_\alpha(E)$ is the retarded self-energy determining the escape rates for electrons to exit into the attached leads, ${\bm \Sigma}^<_\alpha(E)=-f_\alpha[{\bm \Sigma}^r_\alpha(E-eV_\alpha) - {\bm \Sigma}^a_\alpha(E-eV_\alpha)]$ is the corresponding lesser self-energy matrix due to the coupling to the leads, and advanced quantities are defined by ${\bf O}^a = [{\bf O}^r]^\dagger$. In the ``momentum-conserving'' model of dephasing, the interacting self-energies are given by \mbox{${\bm \Sigma}^r_{\rm int}(E)= d_p {\bf G}^r(E)$} and \mbox{${\bm \Sigma}^<_{\rm int}(E)= d_p {\bf G}^<(E)$}, while in the ``momentum-relaxing'' model \mbox{${\bm \Sigma}^r_{\rm int}(E)= \mathcal{D}[d_m {\bf G}^r(E)]$} and \mbox{${\bm \Sigma}^<_{\rm int}(E)= \mathcal{D}[d_m {\bf G}^<(E)]$}~\cite{Golizadeh-Mojarad2007a}. The operator $\mathcal{D}[\ldots]$ selects the diagonal elements of the matrix on which it acts while setting to zero all the off-diagonal elements. Any linear combination of these two choices can be used to adjust the phase and momentum relaxation lengths independently. When computed self-consistently together with ${\bf G}^r(E)$ and ${\bf G}^<(E)$, both of these choices for ${\bm \Sigma}^r_{\rm int}(E)$ and ${\bm \Sigma}^<_{\rm int}(E)$  ensure the conservation of charge current, $\sum_\alpha I_\alpha=0$.

The ``momentum-relaxing'' model of dephasing we adopt here  accounts for {\em local} simultaneous phase and momentum relaxation, and it can be interpreted as a highly simplified version (valid in the high-temperature limit) of the so-called self-consistent Born approximation~\cite{Haug2007} for electron-phonon interaction. This model has also been employed before to study dephasing effects in the integer QHE~\cite{Cresti2008} where phenomenological dephasing length is often employed~\cite{Pryadko1999} to account for electron-electron and electron-phonon interactions without delving into microscopic details.

Assuming that dephasing is localized within the active region of the graphene Hall bar, the spin-resolved charge current in lead $\alpha$ is given by the Meir-Wingreen formula~\cite{Haug2007}
\begin{equation} \label{eq:mw}
I_\alpha^\sigma  =  \frac{e}{h} \int \!\! dE \, {\rm Tr} \, [{\bm \Sigma}_\alpha^{<,\sigma\sigma}(E) {\bf G}^{>,\sigma\sigma}(E) - {\bm \Sigma}^{>,\sigma\sigma}_\alpha(E) {\bf G}^{<,\sigma\sigma}(E)].
\end{equation}
The total charge current in lead $\alpha$ is $I_\alpha = I^\uparrow_\alpha + I^\downarrow_\alpha$ and the total spin current is $I^S_\alpha = I^\uparrow_\alpha - I^\downarrow_\alpha$. The first term in Eq.~\eqref{eq:mw} gives the current flowing from lead $\alpha$ towards the active region (because it is proportional to ${\bf G}^>(E)$ which describes the empty states in the active region), while the second term gives the current flowing in the opposite direction (because it is proportional to ${\bf G}^<(E)$ which describes the occupied states in the active region). Likewise, the self-energies ${\bm \Sigma}^\lessgtr(E)$ are proportional to the occupied lead states and the empty lead states, respectively.

While Eq.~\eqref{eq:mw} is valid both in the linear and non-linear transport regimes, in its original form it is not useful for the analysis of currents and voltages in multiterminal devices. That is, instead of voltages hidden in the self-energy and GF matrices, one would like to recast Eq.~\eqref{eq:mw} into the form similar to Eq.~\eqref{eq:mlb} where one can easily invert such equations to obtain voltages measured between the terminals for known currents injected into the device. For this purpose, we expand all quantities in Eq.~\eqref{eq:negf1} and ~\eqref{eq:negf2} to linear order in $V_\alpha$: ({\em i}) ${\bf G}^r(E) \approx {\bf G}^r_0(E) + {\bf G}^r_0(E)[\sum_\alpha {\bm \Sigma}^r_\alpha(E-eV_\alpha) -{\bm \Sigma}^r_\alpha(E)]{\bf G}^r_0(E)$; ({\em ii}) ${\bm \Sigma}^r_\alpha(E-eV_\alpha) \approx {\bm \Sigma}^{r}_\alpha(E) - e V_\alpha \partial {\bm \Sigma}^{r}_\alpha(E)/\partial E$; and (iii) $f_\alpha(E) \approx f(E) - eV_\alpha \partial f/\partial E$. Here ${\bf G}^r_0(E) = [E- {\bf H} - {\bm \Sigma}^{r}_\alpha(E)]^{-1}$ is the retarded GF in equilibrium, $V_\alpha=\text{const}$. Using this in Eq.~\eqref{eq:mw}, together with the expressions for ${\bm \Sigma}^r_{\rm int}(E)$ and ${\bm \Sigma}^<_{\rm int}(E)$ for ``momentum-relaxing'' dephasing, yields the following generalization of Eq.~\eqref{eq:mlb}
\begin{equation} \label{eq:mld}
I_{\alpha}=\frac{e^2}{h} \sum_\beta (T_{\alpha\beta}^{\rm coh}+T_{\alpha\beta}^{\rm incoh})(V_{\alpha}-V_{\beta}).
\end{equation}
The coherent transmission coefficient is $T_{\alpha\beta}^{\rm coh}=\operatorname{Tr}\, \{ {\bf \Gamma}
_{\alpha}{\bf G}^{r}_0{\bf \Gamma}_{\beta}{\bf G}^{a}_0 \}$, and the incoherent contribution is $T_{\alpha\beta}^{\rm incoh}=\operatorname{Tr}\, \{ {\bf \Gamma}_{\alpha}{\bf G}^{r}_0{\bf \Gamma}_{\beta}^{d}{\bf G}_0^{a} \}$. Using the notation $[{\bf A}]_{jv}$ for the matrix element of ${\bf A}$, the diagonal elements $[{\bf \Gamma}_{\beta}^{d}]_{jj} = d_m \sum_v [{\bf Q}]_{jv} [{\bf G}^{r}_0{\bf \Gamma}_{\beta}{\bf G}_0^{a}]_{vv}$ are expressed~\cite{Cresti2008} in terms of ${\bf Q}=[1-d_m {\bf P}]^{-1}$ and $[{\bf P}]_{jv}=[{\bf G}_0^{r}]_{jv} [{\bf G}_0^{a}]_{vj}$.

In conclusion, we have developed a fully quantum transport theory of recently observed~\cite{Abanin2011} nonlocal voltage in magnetotransport near the DP in graphene Hall bars which provides a unified picture of this phenomenon and the underlying ZSHE from the phase-coherent transport regime at low temperatures to semiclassical transport regime at higher temperatures while allowing one to take into account arbitrary strength of magnetic field or scattering processes near the DP. Our theory starts from the NEGF-based Meir-Wingreen formula, including phenomenological many-body self-energies that take into account relaxation of both the phase and the momentum of Dirac fermions in the active region of the device, which is then linearized to provide connection between current and voltages in different leads thereby generalizing the usual LB formulas for phase-coherent transport in multiterminal geometries.

\begin{acknowledgments}
We thank K. Ensslin for illuminating discussions. C.-L. C. and B. K. N. were supported by DOE Grant No. DE-FG02-07ER46374 and C.-R. C. was supported by Republic of China National Science Council Grant No.  NSC 98-2112-M-002-012-MY3.

\end{acknowledgments}



\end{document}